\documentclass[12pt]{iopart}
\usepackage{graphicx}
\usepackage{dcolumn}
\usepackage{bm}

\usepackage{color}
             
\begin{document}
\title[Mixed phase transition 
and mass-radius constraints of neutron stars]
{Mixed phase transition from hypernuclear matter to deconfined quark matter fulfilling mass-radius constraints 
of neutron stars}

\author{M. Shahrbaf$^{1,2}$, D. Blaschke$^{2,3,4}$ and S. Khanmohamadi$^{1}$}

\address{$^1$ Department of Physics, University of Tehran, P.O.B 14395-547, Tehran, Iran} 
\address{$^2$ Institute for Theoretical Physics, University of Wroclaw, Max Born Pl. 9, 50-204 Wroclaw, Poland}
\address{$^3$ Bogoliubov Laboratory of Theoretical Physics, Joint Institute for Nuclear Research, Joliot-Curie Street 6, 141980 Dubna, Russia}
\address{$^4$ National Research Nuclear University (MEPhI), Kashirskoe Shosse 31, 115409 Moscow, Russia}
\ead{m.shahrbaf46@gmail.com}

\date{\today}

\begin{abstract}
A recent solution of the hyperon puzzle by a first order phase transition to color superconducting quark matter is revisited in order to replace the Maxwell construction by an interpolation method which describes a mixed phase.
To do this, we apply for the first time the finite-range polynomial interpolation method for constructing a transition between hadronic and quark matter phases to the situation that is characterized in the literature as the reconfinement problem.
For the description of the hadronic phase the lowest order constrained variational method is used while for the quark phase the nonlocal Nambu-Jona-Lasinio model with constant (model nlNJLA) and with density-dependent (model nlNJLB) parameters is employed. 
Applying the replacement interpolation method to both quark matter models results in a hybrid equation of state that allows a coexistence of nuclear matter, hypernuclear matter and quark matter in a mixed phase between the pure hadronic and quark phases which can also be realized in the structure of the corresponding hybrid star sequences. 
The predicted hybrid stars fulfill the constraints on the mass-radius relation for neutron stars obtained from recent observations.   
\end{abstract}

\maketitle


\section{Introduction}
While the properties of hot and not too dense, strongly interacting matter are explored in heavy ion collision (HIC) experiments at highest collision energies as well as in lattice quantum chromodynamics (QCD) simulations, the interiors of neutron stars (NS) are the best laboratories in nature where different phases of cold and very dense matter can be investigated \cite{Haensel:2007yy, Lattimer:2012nd}. 
However, NS  can also show features of hot and dense matter, when they are born as proto-neutron stars in a supernova explosion or at the end of their life, when they produce a kilonova in a NS merger event \cite{Rezzolla:2018jee}.

It has been shown that the density of the core of a heavy NS can reach to around 1~fm$^{-3}$ \cite{Haensel:2007yy}. 
Therefore, the Fermi energy level of particles rises sufficiently at such high densities in the inner core of NS that the appearance of the new degrees of freedom is possible \cite{Glendenning:2001pe, Baym:1985tn}. 
Indeed, according to the QCD phase diagram, when the density of matter exceeds the nuclear saturation density ($\rho > \rho_0 \approx 0.16$ fm$^{-3}$), 
in addition to usual nuclear matter, at first the hyperonic degree of freedom may appear. 
At higher densities, a phase transition from baryonic matter to deconfined quark matter is expected and thus, a heavy NS may be a hybrid star \cite{Chen:2012zx}. 
The discovery of pulsars with masses as high as $2~M_\odot$ \cite{Demorest:2010bx, Antoniadis:2013pzd, Lynch:2012vv, vanKerkwijk:2010mt, Fonseca:2016tux, Arzoumanian:2017puf,Cromartie:2019kug} and recently also the observation of the merging of two NS in the event GW170817 \cite{TheLIGOScientific:2017qsa}
has revived this question about the real composition of the core of massive NS \cite{Fischer:2017lag,Bauswein:2018bma}.

Considering hyperons in the core of NS, one faces the hyperon puzzle\footnote{The hyperon puzzle consists in the fact that within Brueckner-Bethe-Goldstone calculations under conservative assumptions for the forces involving hyperons the appearance of hyperons softens the EoS and lowers the maximum mass of a neutron star so that even the well-constrained binary radio pulsar masses of typically $\sim 1.4~M_\odot$ cannot be reached \cite{Baldo:1999rq}. 
It is also present for the NN potential-based LOCVY approach used here.
In other models for hypernuclear matter like, e.g., relativistic mean field models \cite{Hofmann:2000mc,Lastowiecki:2011hh,Maslov:2015msa} or the quark-meson-coupling model
\cite{RikovskaStone:2006ta} it does  not appear. These models can describe neutron stars with a maximum mass  $\sim 2.0~M_\odot$, see also \cite{Blaschke:2018mqw}.}
and its suggested solution by taking in to account the quark deconfinement phase transition from a soft hyperonic equation of state (EoS) to a sufficiently stiff quark matter EoS in the NS interior \cite{Baldo:2003vx,Lastowiecki:2011hh, Shahrbaf:2019vtf}. 
While ab-initio simulations of QCD on a lattice have provided reliable results for the deconfinement transition at finite temperature and zero baryon density 
\cite{Karsch:2000ps, Karsch:2008fe,Borsanyi:2013bia,Bazavov:2014pvz}, such calculations at low temperatures and high baryon densities face the still unsolved sign problem.
Unfortunately, there is not yet a unified theory which could be applied to both hadronic and quark phase in all ranges of densities and temperatures. 
Therefore, it is still an acceptable strategy to calculate the EoS of the hadronic phase and of the quark phase from different reliable theories and then to construct a phase transition between them.

Right after the phase transition, the quark matter EoS at high densities should be sufficiently stiff without violating the causality constraint (the speed of sound should 
not exceed the speed of light, $c_s < 1$) for solving the hyperon puzzle. 
This goal could be achieved using constant speed of sound (CSS) parametrization \cite{Alford:2013aca, Zdunik:2012dj}  for the quark matter EoS at high densities, as in our previous work \cite{Shahrbaf:2019vtf} which hereafter we will refer to as paper I.

While speaking about a strong phase transition, an interesting topic is different methods for constructing a phase transition (PT) from hadronic matter to quark matter which both consist of strongly interacting particles in the NS interior. 
The PT results in a massive hybrid star ($M \approx 2M_\odot$) only if a soft hadronic EoS would be followed by a stiff quark matter EoS. 
This purpose can be obtained employing a Maxwell, Glendenning or pasta PT construction. 
The Maxwell construction garantees the continuity of pressure and chemical potential across the transition \cite{Baym:1976yu, Buballa:2003qv, Benic:2014jia, Khanmohamadi:2019jky, Shahrbaf:2019vtf}. 
The speed of sound vanishes in the mixed phase while employing Maxwell construction since the energy density jumps at a constant critical pressure.
In the Glendenning construction which is also called Gibbs construction, the situation with several globally conserved charges results in a pressure that is monotonously rising throughout the transition \cite{Glendenning:1992vb, Burgio:2002sn}. 
Both Maxwell and Glendenning construction belong to a first-order PT.

A strong first-order PT is accompanied with a large difference in the densities of the subphases which causes strong gradients at their interface. 
Due to nonvanishing surface tension finite size structures occur which are comparable with bubbles and droplets in the boiling transition of water-vapor conversion. 
In this new mixed phase of matter at the border of the transition, different geometrical forms such as planar, cylindrical and spherical structures may appear which have been dubbed "pasta". 
Such finite-size structures can be produced in NS matter because of the global electric neutrality and the surface tension between hadronic and quark phase. 
The surface tension between two different sub-phases and the effect of Coulomb interaction, including screening, affect the size and shape of these structures \cite{Blaschke:2019tbh}. 
As a result, in the mixed phase, the pressure is no longer constant as in the Maxwell construction. 
It is also not changing  as remarkably as in the Glendenning construction where the surface tension is neglected \cite{Glendenning:1992vb}. 
In mixed phase construction, it is supposed that the surface tension between hadronic phase and quark phase does not exceed a critical value of about $\sigma_c \sim 60$ MeVfm$^{-2}$ \cite{Heiselberg:1992dx, Voskresensky:2002hu}. 
This magnitude is governed by the charge screening effects and for every $\sigma > \sigma_c$, the resulting mixed phase reduces to the Maxwell construction, see
also \cite{Maslov:2018ghi}. 
In particular, for a massive hybrid star ($M > 2~M_\odot$), it is important to consider the effects of finite size and pasta phases at the hadron-quark transition \cite{Yasutake:2014oxa}. Therefore, it seems that the obtained results for NS are more physical and trustworthy when using a mixed PT instead of the Maxwell case. 
A detailed discussion of hadron-quark transition mimicking the pasta phase is provided in the literature; see, e.g., \cite{Yasutake:2014oxa, Spinella:2015ksa, Na:2012td}.

It is hard to consider all details about pasta phase but it has been shown that a simple parabolic interpolation of pressure as a function of chemical potential 
\cite{Ayriyan:2017tvl} has a good agreement with a full pasta phase calculation \cite{Maslov:2018ghi}. 
The method of interpolation between hadron and quark EoS for neutron star application has been suggested for the first time in 2013 \cite{Masuda:2012kf, Masuda:2012ed}. This method has been developed in analogy to another technique applied along the temperature axis in QCD phase diagram \cite{Asakawa:1995zu}
where the baryonic chemical potential is vanishing. 
However, after correcting shortcomings of the first work in the subsequent ones, the revised and corrected version of the interpolation method has been employed extensively instead of a reliable unified approach which covers all ranges of density and temperature \cite{Kojo:2014rca, Blaschke:2013rma}. 
Therefore, for describing the cold and dense matter which is our focus in this work, at low and high densities we employ the hadronic EoS and quark EoS respectively while in the intermediate regime, where there is not enough tools and knowledge to perform reliable calculations, an interpolated EoS between two phases will be employed \cite{Kojo:2014rca, Kojo:2015fua, Kojo:2015nzn}.
We like to remark that these interpolation methods are using the $\tanh$-function to switch from one phase to the other, so that strictly speaking there are effects of the quark matter phase on low-density hadronic matter and nuclear matter effects on high-density quark matter, because the choice of this mathematical function does not restrict the mixed phase to a finite domain bounded by the limiting densities of the applicability of pure hadronic and quark matter models, respectively.  
This caveat has been removed by the recently developed replacement interpolation method (RIM) \cite{Ayriyan:2017tvl,Ayriyan:2017nby} and by the mixing interpolation method (MIM) \cite{Alvarez-Castillo:2018rrv}. Both approaches are explained and compared with each other, together with their applications to NS physics in Ref.~\cite{Abgaryan:2018gqp}. 

Following  paper I, in which a Maxwell construction has been employed to describe hybrid stars with hypernuclear and superconducting quark matter in their interior, 
we intend in this work to improve our calculations by considering mixed phases in the transition region between hadronic and quark phase applying the RIM. 
It is worth mentioning that the low-density and high-density EoS are not trustful right before and after the crossing point since a soft hyperonic EoS is extrapolating to high densities where the quark exchange should be dominant. On the other hand, a perturbative stiff EoS is employed in to the low-density region around the nuclear saturation density while it doesn't include confinement effects. Furthermore, the low density region around nuclear matter at zero temperature is subject to chiral symmetry and thus not compatible with the quark matter EoS.

The motivation of this work is to investigate how a mixed PT from hypernuclear matter to color superconducting quark matter can improve the results and the properties of hybrid star. 
In order to do this, we compute numerically the EoS of hybrid star matter mimicking pasta structures using the lowest order constrained variational method 
\cite{Owen:1977uun} for the hypernuclear matter EoS and the nonlocal Nambu-Jona-Lasinio model \cite{GomezDumm:2005hy, Blaschke:2007ri} for deconfined, 
color superconducting quark matter. 
Both of them have been employed widely for describing the NS properties. 
As it has been mentioned in paper I, two schemes have been used for nonlocal Nambu-Jona-Lasinio (nlNJL) model in which our approach of treating with the values of the coupling constants is different. In "model nlNJLA", we consider a set of coupling constants to be fixed at finite densities. The second one is a generalization of model nlNJLA to the case when parameters of the model (such as coupling constants) become functions of the baryochemical potential \cite{Alvarez-Castillo:2018pve}; we denote this case as "model nlNJLB". 
Our motivation for this work is to improve the results for the PT in both models by using the RIM between hyperonic EoS and quark matter EoS.

In Sec. II we present a brief description of the lowest order constrained variational (LOCV) method as well as the nlNJL model. The Sec. III is devoted to the introduction of the RIM for constructing the 
mixed PT and in Sec. IV the results of this calculation are presented and discussed. 
Sec. V is devoted to the properties of the resulting hybrid stars.  Finally, the summary and conclusion are given in Sec. V.
\section{State-of-the-art equations of state for hypernuclear and quark matter}
For anticipating the properties of hybrid stars at zero temperature, trustful EoS for both hadronic and quark phase are essential.
We have started a research that joins two different domains of state-of-the-art expertise: the lowest order constrained variational method with hyperons (LOCVY) for hypernuclear matter and the color superconducting chiral nlNJL model for quark matter.
The theoretical approaches used to calculate the EoS for each phase will be briefly discussed in the following two subsections.
\subsection{Hadronic phase: Hypernuclear matter within the LOCVY method}
A microscopic interaction-based variational method called LOCV method is employed for the nuclear matter phase. As for all variational methods, the starting point of the calculations is the Hamiltonian of the nuclear many-body system
\begin{equation}
H=\sum _{i}\frac{p_{i} ^{2} }{2m}  +\sum _{i<j}V(ij),
\end{equation}
where the momentum of $i$th particle is $p_{i}=-\hbar\nabla_{i}$, and $V(ij)$ is the realistic two-body potential for every pair of particles \cite{Bishop:1978ix}. This method has been employed within the years for calculating the bulk properties of nuclear matter \cite{Modarres:2000nk, Moshfegh:2005rom,Moshfegh:2007mxh} as well as the mass-radius relations of NSs \cite{Goudarzi:2015dax}, hyperonic stars \cite{Shahrbaf:2019wex} and hybrid stars \cite{Shahrbaf:2019vtf}. It has been recently developed to calculate the energy per particle and correlation functions in hypernuclear matter \cite{Shahrbaf:2019bef}. For nucleon-nucleon interaction, the AV18 potential \cite{Wiringa:1994wb} supplemented by Urbana type three-body force \cite{Goudarzi:2015dax} has been employed while for nucleon-hyperon and hyperon-hyperon interactions, we have used a three-ranges Gaussian potential proposed by Hiyama et al. \cite{Hiyama:2006xv,Hiyama:2002yj}. 
The trial wave function for $A$-body interacting systems then reads
\begin{equation}
\Psi(1...A)=F(1...A)\Phi(1...A),
\end{equation}
where $\Phi$ is the uncorrelated wave function, i.e., the Slater determinant of one-body ideal Fermi gas wave functions in $A$-body system and $F$ is the correlation function of $A$-body system which is written in Jastrow form \cite{Shahrbaf:2019wex}. 
In agreement with other variational method such as APR \cite{Akmal:1998cf}, a set of Euler - Lagrange differential equations is obtained by minimizing the energy with respect to the correlation function under a constraint. The correlation functions and subsequently the energy of the system per particle is calculated by solving these equations \cite{Bordbar:1998xv}.\\
In the LOCV method, the only constraint used is normalization condition described by
\begin{equation}
 \label{eq:1}
 \chi=\left\langle \Psi|\Psi\right\rangle - 1
=\frac{1}{A} \sum _{ij}\langle ij|F_{p}^{2}-f^{2}|ij-ji\rangle  =0,
\end{equation}
where $F_{p}$ is the Pauli function, $f$ denotes to the two-body state-dependent correlation functions, $\langle ij|$ is the two-particle eigenstate which includes the total angular momentum, orbital angular momentum, spin, isospin, the third component of isospin and the strangeness number of two particles. 
$|ij-ji \rangle$ is considered to satisfy the anti-symmetry properties of fermionic states under particle interchange.\\
For more details about the LOCVY method calculations, one can read \cite{Shahrbaf:2019bef, Shahrbaf:2019vtf} and the references therein.
\subsection{Quark matter EoS within nlNJL model}
For the quark matter phase we employ a color superconducting nonlocal chiral quark model of the Nambu-Jona-Lasinio type (nlNJL) for the case of two quark flavors. This method is a nonlocal covariant extension of the NJL model which is one of the most popular methods among the chiral effective models. In NJL model, the quark fields interact via local four-point vertices which are subject to chiral symmetry characterized by four-fermion interactions which are nonlocal. A problem with the NJL model, related to the use of local interactions, is that some type of regularization must be introduced to cure divergent integrals, introducing certain ambiguity in the choice of the regularization scheme. In nlNJL model, the dynamical mass function $M(p)$ which shows the momentum dependence of the quark mass as well as the momentum current which is responsible for a momentum dependent wave function renormalization (WFR), $Z(P)$ of the quark propagator in the vacuum are introduced which are fitted to lattice QCD data \cite{Contrera:2016rqj}. This model has recently been discussed in the context of the third family of compact stars \cite{Alvarez-Castillo:2018pve}, where also references to preceding works are given.\\
The nlNJL approach is characterized by four-fermion interactions in the scalar quark-antiquark, the anti-triplet scalar diquark and the vector quark-antiquark channels.
The effective Euclidean action for two light flavors is written as below
\begin{eqnarray}
S_E &=& \int d^4 x \ \left\{ \bar \psi (x) \left(- i \rlap/\partial + m_c
\right) \psi (x) - \frac{G_S}{2} j^f_S(x) j^f_S(x) 
\right.\nonumber\\
&& \left. 
- \frac{H}{2}
\left[j^a_D(x)\right]^\dagger j^a_D(x) {-}
\frac{G_V}{2} j_V^{\mu}(x)\, j_V^{{\mu}}(x) \right\} \, 
\label{action}
\end{eqnarray}
where $j_{S}(x)$, $j_{D}(x)$ and $j_{V}(x)$ are the scalar, diquark and vector currents of quarks, respectively.
$G_S$ is the scalar coupling strength, while $H/G_S$ and $\eta=G_V/G_S$ are the coupling constant ratios in the diquark and vector channels, resp., 
normalized to the scalar coupling.
It is supposed that the current quark mass, $m_c$ is the same for $u$ and $d$ quarks. 
One should note that the currents $j_{S,D,V}(x)$ are nonlocal and based on a separable approximation of the effective one gluon exchange (OGE) model of QCD. The input parameters are the coupling constants ratios $H/G_S$ and $\eta$.
However, it has been estimated from Fierz transformation for OGE interactions in the vacuum that $H/G_S =0.75$ and $\eta= 0.5$, but there is no powerful phenomenological constraint on the ratio $H/G_S$ and the values are subject to large theoretical uncertainties \cite{GomezDumm:2005hy}. Because of these uncertainties, we have considered two schemes of fixing the values of coupling constant in paper I which have been introduced in previous section.\\ 
In mean field approximation (MFA), the grand canonical thermodynamic potential per unit volume reads
\begin{eqnarray}
\Omega^{\rm MF}  &=&   \frac{ \bar
\sigma^2 }{2 G_S} + \frac{ {\bar \Delta}^2}{2 H} 
- \frac{\bar \omega^2}{2 G_V} \nonumber\\
&&- \frac{1}{2} \int \frac{d^4 p}{(2\pi)^4} \ \ln
\mbox{det} \left[ \ S^{-1}(\{\mu_{fc}\};\bar \sigma, \bar \Delta, \bar \omega) \right] \ . \label{mfaqmtp}
\end{eqnarray}
where $\{\mu_{fc}\}$ is the set of chemical potentials for quarks of flavor $f$ and color $c$, while $\bar \sigma$, $\bar \Delta$, $\bar \omega$ are the mean field values in the scalar meson, diquark and vector meson channels, respectively.
More details about the calculation of the explicit expression for the thermodynamic
potential can be found in \cite{Blaschke:2007ri}.
The following coupled equations are satisfied by the mean field values $\bar \sigma$, $\bar \Delta$ and $\bar \omega$
\begin{eqnarray}
\frac{ d \Omega^{\rm MF} }{d\bar \sigma} \ = \ 0 \ , \ \ \
\frac{ d \Omega^{\rm MF} }{d\bar \Delta} \ = \ 0 \ , \ \ \
\frac{ d \Omega^{\rm MF} }{d\bar \omega} \ = \ 0 \ .
\label{gapeq}
\end{eqnarray}
The chemical potentials of quarks in chemical equilibrium are related to each other by considering the baryonic chemical potential $\mu$, a quark electric chemical potential $\mu_{Q_q}$ and a color chemical potential $\mu_8$ as the Lagrange multipliers which adjust the conserved baryon number, electric charge and color charge of the system. 
The chemical potential $\mu_{Q_q}$ distinguishes between up and down quarks, and the color chemical potential $\mu_8$ has to be introduced to ensure color neutrality.
For applying this quark matter scheme in NS applications, one has to consider the $\beta$-equilibrium conditions. 
We shall consider that quark matter is electrically and color neutral and in equilibrium under weak interactions. 
By self-consistently solving the gap equations (\ref{gapeq}) along with $\beta-$ equilibrium conditions for chemical potentials as well as electric and color charge neutrality conditions, one obtains $\bar \Delta$, $\bar \sigma$, $\mu_l$ and $\mu_8$ for each value of $\mu$. It is recommended to read the  Appendix of \cite{Blaschke:2007ri} as well.

The quark matter EoS is then
\begin{equation}
\label{eq:P-mu}
P(\mu)=P(\mu;\eta(\mu),B(\mu))=-\Omega^{\rm MF}(\eta(\mu)) - B(\mu)~,
\end{equation}
in which the vector coupling strength parameter $\eta(\mu)$ and the bag pressure $B(\mu)$ are here considered as functions of the baryon chemical potential. 
For more details, see \cite{Alvarez-Castillo:2018pve,Blaschke:2020qqj}.
As it has been mentioned in paper I, the medium dependence of the gluon sector is considered in the bag pressure $B(\mu)$. As a result, in quark matter EoS of model nlNJLB, both parameters $\eta$ and $B$ may depend on the chemical potential while in quark matter EoS of model nlNJLA, they are supposed to be constant. 
In paper I, we have employed both schemes for quark matter along with the LOCVY method in the context of the new constraints of NS \cite{Shahrbaf:2019vtf}.
%
\section{Phase transition construction}
As it was described in previous sections, there is no unified EoS which can be applied to all ranges of densities and temperatures. It is physical to employ the hadronic EoS at low densities around nuclear saturation density where the mixed structure of the baryons has not appeared. According to the QCD phase diagram which is progressing by different methods and will be probed in the future heavy ion collision experiments such as FAIR and NICA, considering hyperons in nuclear matter can slightly increase the density of the validity of hadronic. We define the upper limit of the validity of our hyperonic EoS as $n_H(\mu)$.\\
At relatively high densities, where the quarks don't belong to a specific baryon any more, the deconfined quark matter EoS has to be employed. The lower limit of the validity of the color superconducting quark matter EoS is denoted by $n_Q(\mu)$. At the intermediate range of density, i.e., $n_H(\mu) < n(\mu) < n_Q(\mu)$, neither hyperonic EoS nor quark matter EoS are applicable.

Since the EoS at the the intermediate range of density is not well defined, it is assumed that the EoS of both phases changes due to the finite size and Coulomb effects around the phase transition point which can be obtained using the Maxwell construction. Furthermore, the surface tension between the hadronic phase and quark phase in strongly interacting system is not well known. If the value of surface tension is infinite, a Maxwell construction can be used for phase transition while a vanishing surface tension corresponds to the Glendenning construction. 
For mixed phase transition which mimics the pasta phase, we assume a variable surface tension.\\
 The exact description of the physics of pasta phase is not straightforward because it requires the size and shape of structures as well as transitions between them to be taken in to account but it has been studied in different works by different methods \cite{Voskresensky:2002hu, Yasutake:2014oxa, Maruyama:2007ss, Watanabe:2003xu, Horowitz:2005zb, Horowitz:2014xca, Newton:2009zz}.
 
In this work, we use the replacement interpolation method (RIM) \cite{Abgaryan:2018gqp} in which a simple modification of the Maxwell construction is employed instead of a full solution of pasta phase. Since the EoS of hyperonic and quark matter phases are described with the relation between the pressure and chemical potential (for $T = 0$ which is applicable for the NS), $P_H(\mu)$ and $P_Q(\mu)$ respectively, the effective mixed phase EoS $P_M(\mu)$ can be described simply by an interpolated function between two phases. It requires that the interpolated pressure coincides the hyperonic and quark values at lower and upper limits along with satisfying the thermodynamic constraint of positive slope of density versus chemical potential, i.e., $\frac{\partial n_M}{\partial \mu_M} = \frac{\partial^2 P_M}{\partial \mu_M^2}  > 0$, as well as the causality condition that the adiabatic speed of sound at zero frequency, $c_s^2 = \frac{\partial P}{\partial \epsilon}$, does not exceed the speed of light. A simple and reasonable function to interpolate the pressure is a polynomial function which smoothly joins the pressure curves of two phases.
This method has been developed in \cite{Ayriyan:2017tvl} and applied to the question of robustness of NS mass twins against mixed phase effects in 
\cite{Ayriyan:2017nby}. We repeat here the basic steps of its derivation following Refs.~\cite{Ayriyan:2017tvl,Ayriyan:2017nby}.

The value of the critical baryochemical potential $\mu_c$ for which the phases are in mechanical and chemical equilibrium with each other is obtained from the Maxwell condition
  \begin{equation}
  P_Q(\mu_c)~=~P_H(\mu_c)~=~P_c~.
   \label{eq:mechequi}
  \end{equation}
For the pressure of the mixed phase a polynomial ansatz is considered
\begin{equation}
 P_M(\mu)~=~\sum_{q=1}^{N}\alpha_q(\mu-\mu_c)^q~+~(1+\Delta_P)P_c,
  \label{eq:RIMpressure}
\end{equation}
in which $\Delta_P$ is a free parameter which determines the pressure of mixed phase at $\mu_c$
\begin{equation}
 P_M(\mu_c)~=~P_c+\Delta_P~=~P_M~,~\Delta_P=\Delta P/P_c.
  \label{eq:addpressure}
\end{equation}
The value of $\Delta_P$ is related to the surface tension between two phases such that the vanishing $\Delta_P$ corresponds to a minimal value of the surface tension for which the transition becomes equivalent to that of the Maxwell construction in which the pressure at the critical point is constant.
The quantitative relation between $\Delta_P$ and the surface tension has been given in Ref.~\cite{Maslov:2018ghi}
for a selection of hybrid EoS cases and it shows that a value of $\Delta_P \approx 0.05-0.07$ corresponds to a vanishing surface tension and thus a Glendenning construction \cite{Glendenning:1992vb}. 

Generally in (\ref{eq:RIMpressure}), one can consider \cite{Abgaryan:2018gqp}
\begin{equation}
 N~=~2k~,~k=1,2,...
  \label{eq:nnnn}
\end{equation}
According to the Gibbs conditions for phase equilibrium (\ref{eq:mechequi}) at the matching points  $\mu_{H}$ and $\mu_{Q}$ of the mixed phase pressure with the pressure of hyperonic and quark matter EoS, the pressures and their derivative of order $k$ have to satisfy the continuity conditions
\begin{eqnarray}
P_H(\mu_{H}) &=&  P_M(\mu_{H})~,\\
P_Q(\mu_{Q}) &=&  P_M(\mu_{Q})~,\\
\frac{\partial^k}{\partial\mu^k}P_H(\mu_{H}) &=&  \frac{\partial^k}{\partial\mu^k}P_M(\mu_{H})~,\\
\frac{\partial^k}{\partial\mu^k}P_Q(\mu_{Q}) &=&  \frac{\partial^k}{\partial\mu^k}P_M(\mu_{Q})~.
\label{eq:continuity}
\end{eqnarray}
For ease of calculation, we assume that the effective mixed phase pressure could be described in the parabolic form
\begin{equation}
 P_M(\mu)~=~\alpha_2(\mu-\mu_c)^2~+~\alpha_1(\mu-\mu_c)~+~(1+\Delta_P)P_c,
  \label{eq:pppressure}
\end{equation}
for which the $\alpha_1$, $\alpha_2$ as well as $\mu_{H}$ and $\mu_{Q}$ could be obtained from the continuity conditions when $k=1$. 
By solving the continuity equations, one can obtain \cite{Ayriyan:2017tvl}
\begin{eqnarray}
\alpha_2 &=& \frac{1}{\mu_{Q}-\mu_{H}}~\left(\frac{P_Q-P_M}{\mu_Q-\mu_c}~-~\frac{P_H-P_M}{\mu_H-\mu_c}\right)~,\\
\alpha_1 &=& \frac{P_M-P_H}{\mu_{c}-\mu_{H}}~+~\frac{P_Q-P_M}{\mu_Q-\mu_c}~-~\frac{P_Q-P_H}{\mu_Q-\mu_H}~.
\label{eq:coefficient}
\end{eqnarray}
Considering $n(\mu)=dP(\mu)/d\mu$, we numerically solve the continuity equations for baryon density and obtain $\mu_{H}$ and $\mu_{Q}$.

It is worth mentioning that this interpolation method can be applied not only to the usual phase transition from the hadronic phase to the quark phase with 
$\Delta_P > 0$ but also to the case where applying the principle of maximum pressure to the crossing of the pressure vs. chemical potential curves for quark and hadron matter would correspond to a nonphysical reconfinement transition \cite{Zdunik:2012dj} from quark phase to hadronic phase with $\Delta_P < 0$. 
In next section, we shall employ the RIM to both these cases which have been constructed in paper I. 
\section{Results and discussion}
Since this study is a follow-up of paper I, we review here some results of that work for the convenience of the reader. 
As it has been reported in paper I, we have constructed a phase transition using the Maxwell construction from the hyperonic EoS obtained by the LOCVY method to a color superconducting quark matter EoS obtained from the nlNJL model. 
For the quark matter EoS, we have used two different approaches denoted as model nlNJLA and model nlNJLB. 
In model nlNJLA with constant parameters, the transition occurs at low densities, even before the appearance of hyperons because of the lack of confining effects like a bag pressure. In this model, at higher densities the unphysical crossing of hadronic and quark matter EoS occurs which would correspond to a reconfinement transition.
However, in the model nlNJLB with density dependent parameters, an intermediate hypernuclear matter phase in the hybrid star, between the nuclear and color superconducting quark matter phases is anticipated while the resulting hybrid star obeys the new constraints of mass-radius of NS.

Our motivation of this work is extending the calculation of paper I to improve the results of model nlNJLA and to obtain a more physical results in both approaches. 
To this end, we first inspect the lower panel of Fig. 1 of paper I. It is clear that the curve for (hyper)nuclear matter pressure crosses the ones for the quark matter twice. The first crossing is the usual transition from the hadronic phase to the quark phase at low densities, but the second one would correspond to a transition from quark to hadronic matter (reconfinement) when increasing the density.

Therefore, we are interested to employ the interpolation method between hadronic and quark phase for two reasons:
\begin{enumerate}
 \item 
 Cure the problem of model nlNJLA by ignoring the Maxwell transition at low densities and consider a mixed phase transition from hadronic to quark matter in the 
 region of reconfinement.
 \item 
 Improve the results of transition in both models, nlNJLA and nlNJLB, by investigating a more physical transition by which a mixture of nuclear, hypernuclear and deconfined quark matter in the transition region as well as more realistic hybrid star sequences are predicted.  
\end{enumerate}

\begin{figure*}[!th]
\centering
\includegraphics[width=0.8\textwidth]{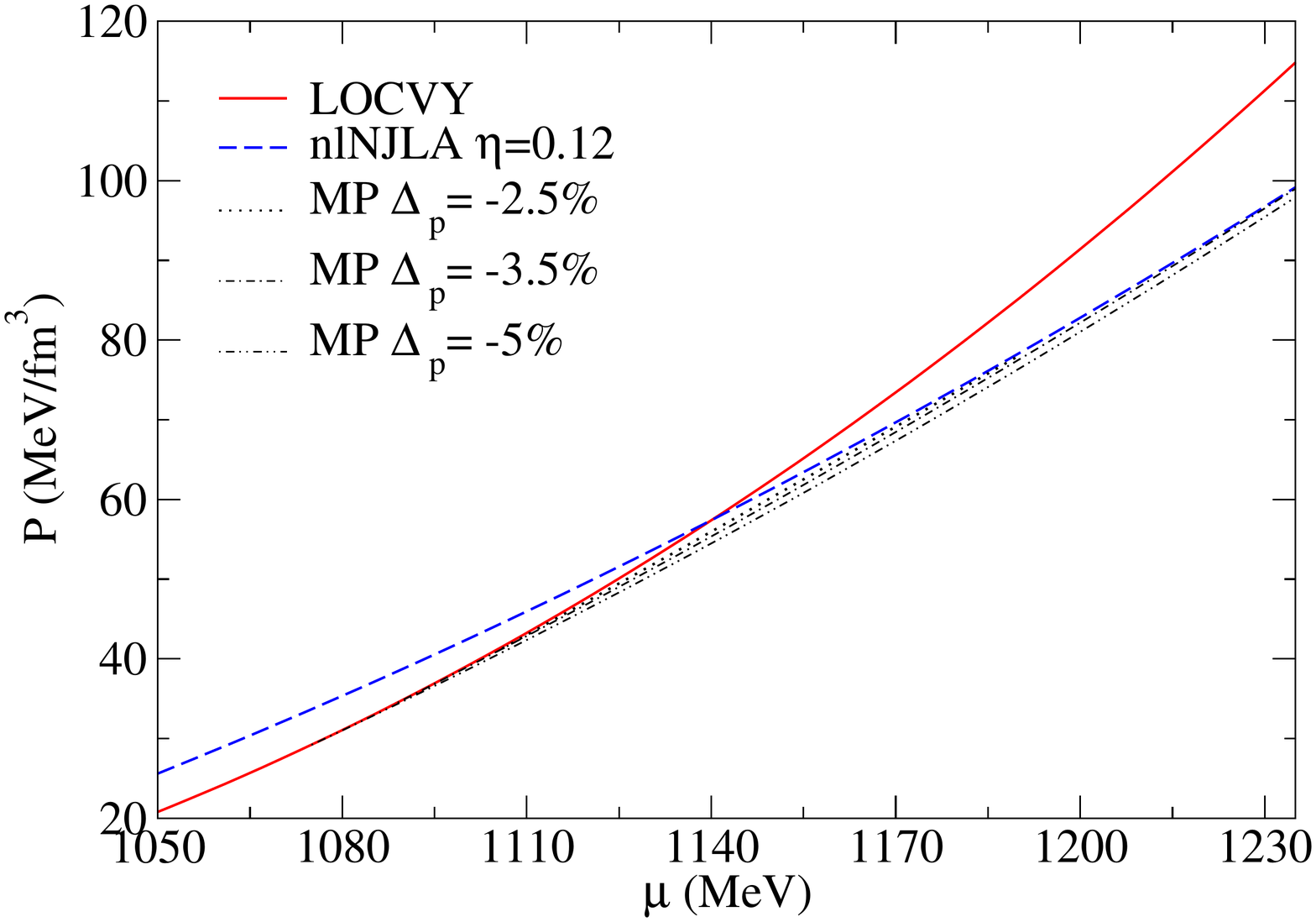}
\caption{\label{fig:1}Pressure as a function of chemical potential for mixed phase transition from hypernuclear matter obtained from LOCVY method to 
color superconducting deconfined quark matter obtained from model nlNJLA for $\eta = 0.12$ and $\Delta_P = -2.5\%, -3.5\%, -5\%$. }
\end{figure*}

At the second crossing point of model nlNJLA which is under the scope of our study in this paper, the hyperonic pressure goes beyond the quark matter pressure which is not physically acceptable at high densities. 
Thus, we have employed the interpolation method with $\Delta_P < 0$ in such a way that a transition from hyperonic matter to deconfined quark matter occurs. Since the basic hypothesis of the interpolation method is changing the hyperonic and quark matter EoS around the transition region, this method makes sense.
It must be remarked, however, that a regular Maxwell construction at this second crossing point would not make sense. 
Consistent with the RIM, we replace the unphysical parts of both EoS in a finite interval around the crossing point by a parabolic (in general, a polynomial) interpolation that fulfils the matching conditions given in the previous section.

 In paper I, we have studied the Maxwell PT in model nlNJLA for four different values of $\eta$. In this paper, we choose $\eta = 0.12$ to do the mixed phase transition numerical calculations. Therefore, we show that even for the value of $\eta$ for which the first crossing point occurs at the lowest chemical potential and the second one occurs at the highest chemical potential, the results of mixed phase are good for three different values of $\Delta_P$. 
 Fig.~\ref{fig:1} shows the EoS for pressure as a function of chemical potential which has been obtained from the interpolation method. 
 Therefore, a mixed phase construction has been applied to this region. 
 The behavior of the hyperonic and quark matter EoSs is clear around the transition region in Fig. \ref{fig:1}. They have changed in a parabolic shape so that the hyperonic EoS is dominant at lower densities while the quark matter EoS is dominant at higher densities. This result is in agreement with what we expect from a physical system.
 
 For model nlNJLB, the nuclear matter connects to deconfined quark matter by a hypernuclear matter shell at the transition point by a Maxwell construction. 
 But the assumption of infinite surface tension in the Maxwell construction which does not allow to different phases to be mixed with each other can be improved by a mixed phase construction in which the parameter of $\Delta_P$ is related to surface tension.
\begin{figure*}[!th]
\centering
\includegraphics[width=0.8\textwidth]{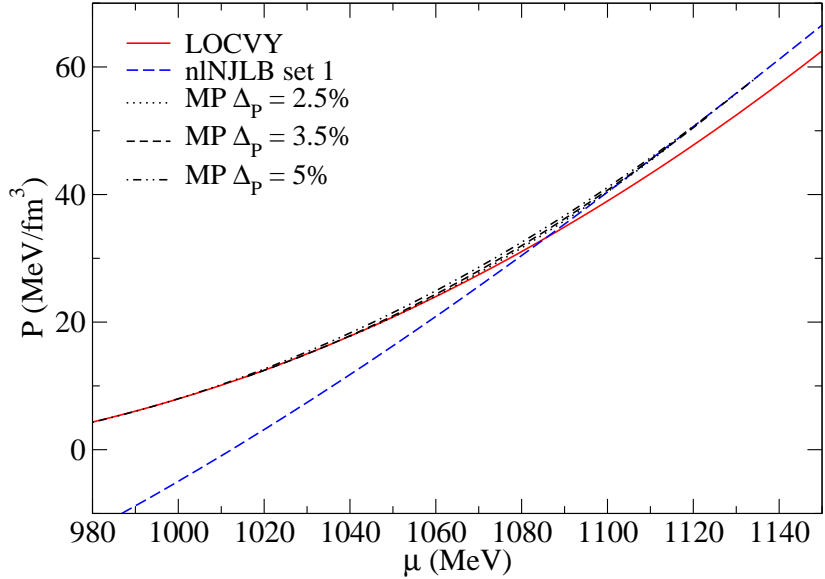}
\caption{\label{fig:2} Pressure as a function of chemical potential for mixed phase transition from hypernuclear matter obtained from LOCVY method to 
color superconducting deconfined quark matter obtained from model nlNJLB - set 1 and $\Delta_P = 2.5\%, 3.5\%, 5\%$.}
\end{figure*}

\begin{figure*}[!ht]
\centering
\includegraphics[width=0.8\textwidth]{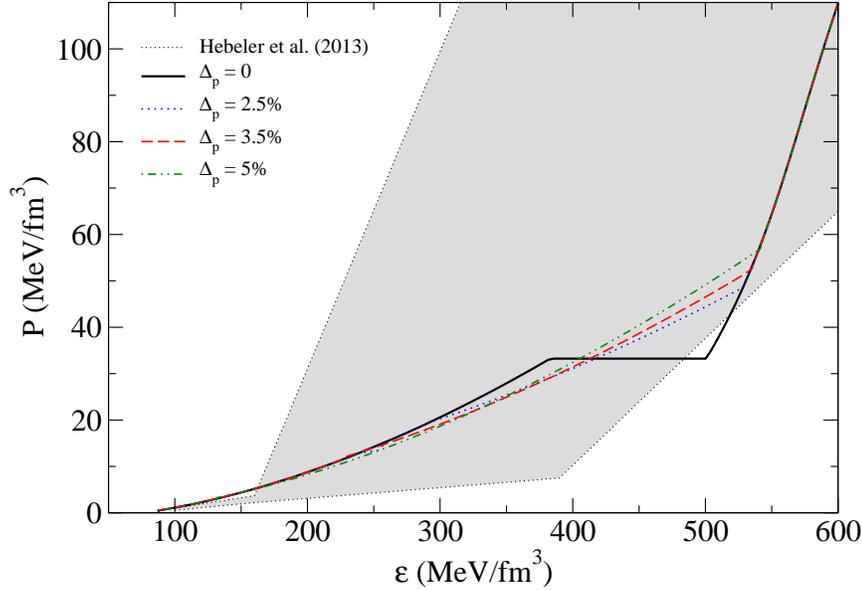}
\caption{\label{fig:3} Pressure as a function of energy density for mixed phase transition from hypernuclear matter obtained from LOCVY method to color 
superconducting deconfined quark matter  obtained from model nlNJLB - set 1 and $\Delta_P = 2.5\%, 3.5\%, 5\%$. 
The hatched region corresponds to the EoS constraint from Ref. \cite{Hebeler:2013nza}.}
\end{figure*}
 
 Therefore, we have employed the same interpolation method with $\Delta_P > 0$ to the phase transition which has been constructed before by the Maxwell construction method for model nlNJLB in paper I. 
 As it has been mentioned in that paper, for model nlNJLB with density dependent parameters like vector meson coupling strength and bag pressure, four different sets of parameters have been used which have been listed in table I of paper I. 
 The continuity condition equations have been numerically solved for the set 1 of model nlNJLB and the results are shown displayed in Fig. \ref{fig:2} and Fig. \ref{fig:3}. The pressure of a hybrid star obtained by a mixed phase construction as a function of chemical potential has been plotted in Fig. \ref{fig:2} while Fig. \ref{fig:3} shows the pressure of the same hybrid star as a function of energy density.
 
 As it can be seen, the threshold of the transition moves to lower density by increasing the value of $\Delta_P$, but since this threshold corresponds to a mixed phase transition, we don't worry about it. 
 At the mixed phase region, by increasing the density (chemical potential), the hypernuclear matter as well as the deconfined quark matter can appear after ordinary nuclear matter. 
In Tab.~\ref{tab:mixed} we summarize the dependence of the chemical potentials and densities at the borders of the mixed phase on the value of the parameter 
$\Delta_P$ for the mixed phase construction of LOCVY hypernuclear matter with nlNJLA and nlNJLB quark matter, respectively.
\begin{table}[h]
 \centering
 \caption{\label{tab:mixed}
  Limiting chemical potentials $\mu_H$ ($\mu_Q$) and density $n_H$ ($n_Q$) on the hadronic phase (quark phase) border of the mixed phase construction 
  for the hybrid EoS using the LOCVY method for the hadronic phase and nlNJLA or nlNJLB model for the quark phase.}
  \vspace{5mm}
 \begin{tabular}{lcccc|cccc}
 model&$\Delta_P$ &$\mu_H$ & $\mu_Q$ & $\mu_H-\mu_Q$ & $n_H$ & $n_Q$ & $n_H-n_Q$ \\
	&  & [MeV] & [MeV] & [MeV] & [1/fm$^3$] & [1/fm$^3$] & [1/fm$^3$] \\
 \hline
\textrm{nlNJLA}&-0.025 & 1106.27& 1189.75& 83.48 & 0.427 & 0.447 & 0.02  \\
 \textrm{nlNJLA}&-0.035 & 1089.46 &1250.94 & 161.48 & 0.392  & 0.502 & 0.11   \\
 \textrm{nlNJLA}&-0.050 & 1075.536 & 1392.68 & 317.144 & 0.362 & 0.631 & 0.269\\ \hline
 \textrm{nlNJLB}&0.025 & 1040.42 & 1115.54 & 75.12 & 0.288  & 0.512 & 0.224   \\
 \textrm{nlNJLB}&0.035 & 1016.88 & 1124.37 & 107.49 & 0.237 & 0.524& 0.287  \\
 \textrm{nlNJLB}&0.050 & 971.41 & 1135.98 & 164.57 & 0.138 & 0.536 & 0.398  \\
 %
 \end{tabular}
 \end{table}

\section{Neutron star properties with mixed phase construction}  
As natural cosmological laboratories, the neutron star properties can verify the applicability of the interpolated EoS. 
A reliable theoretical calculation has to be verified by the observational constraints for the lower limit on the NS maximum mass, currently represented by  the Shapiro-delay measurement for the millisecond pulsar PSR J0740+6620 \cite{Cromartie:2019kug} and by the tidal deformability constraint for the binary NS merger GW170817 \cite{TheLIGOScientific:2017qsa, Bauswein:2017vtn, Annala:2017llu}.
We consider the hybrid star obtained by the mixed phase construction as a hydrostatically equilibrated and spherically symmetric system. 
Thus, the mass-radius relation of the star can be obtained for a given EoS by solving the well-known Tolman-Oppenheimer-Volkoff (TOV) equations  \cite{Tolman:1939jz,Oppenheimer:1939ne}, 
\begin{eqnarray}
\frac{dP(r)}{dr} &=&-\frac{GM(r)\varepsilon (r)}{c^{2} r^{2} } \left(1+\frac{P(r)}{\varepsilon (r)} \right) 
\left(1+\frac{4\pi r^{3} P(r)}{M(r)c^{2} } \right)
 \left(1-\frac{2GM(r)}{rc^{2} } \right)^{-1},\nonumber\\
\frac{dM(r)}{dr} &=&\frac{4\pi \varepsilon (r)r^{2} }{c^{2} }.
\end{eqnarray}
In these equations $P(r)$ and $\varepsilon(r)$ denote the pressure and the energy density profiles for the matter distribution in the NS interior, $M(r)$ is the cumulative mass enclosed in a spherical volume at the distance $r$ from the center, and $G$ is the gravitational constant. The gravitational mass  $M=M(r=R)$  of the star is the mass enclosed within the radius of the star. 
By considering that the boundary condition $P(r=R)=0$, the radius $R$ is defined. 
For a chosen central energy density $\varepsilon_c=\varepsilon(r=0)$, we have the necessary boundary and initial conditions to solve the TOV equations for a given EoS and obtain a relativistic star with mass $M$ and radius $R$, respectively. 
By increasing $\varepsilon$ (or equivalently $P$) at the center of the configuration, the mass-radius relation can be obtained and the maximum mass can be identified from it. 
For the EoS of the inner and outer crust of the neutron star, we use the results of Negele and Vautherin \cite{Negele:1971vb} and Harrison and Wheeler \cite{harrisonwheeler}, respectively.
\begin{figure*}[!th]
\centering
\includegraphics[width=0.8\textwidth]{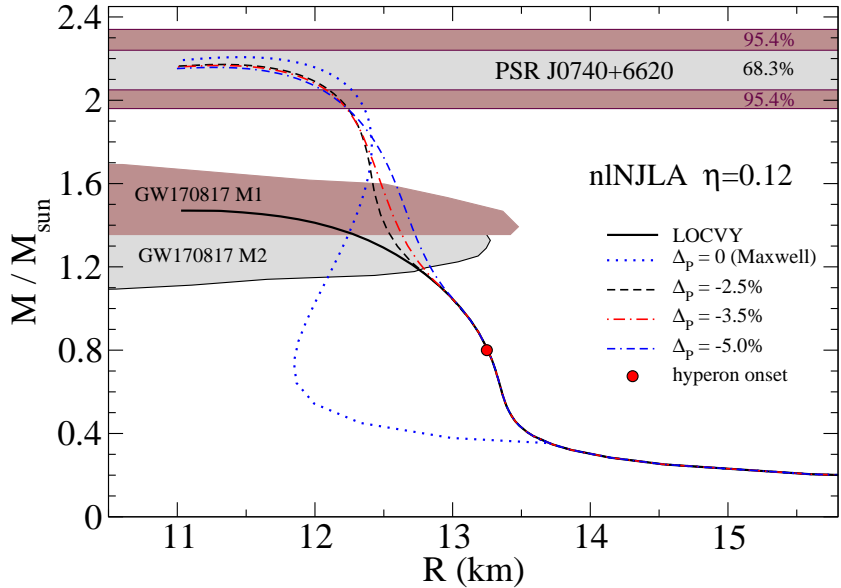}
\caption{\label{fig:4} Mass-radius relation of hybrid star for mixed phase construction of the deconfinement PT using LOCVY method for the hadronic phase and the color superconducting nlNJLA model for quark matter.
The hatched regions correspond to the constraints on the
maximum mass from PSR J0740+6620 \cite{Cromartie:2019kug} and on the compactness from GW170817 \cite{TheLIGOScientific:2017qsa}.}
\end{figure*}

\begin{figure*}[!ht]
\centering
\includegraphics[width=0.8\textwidth]{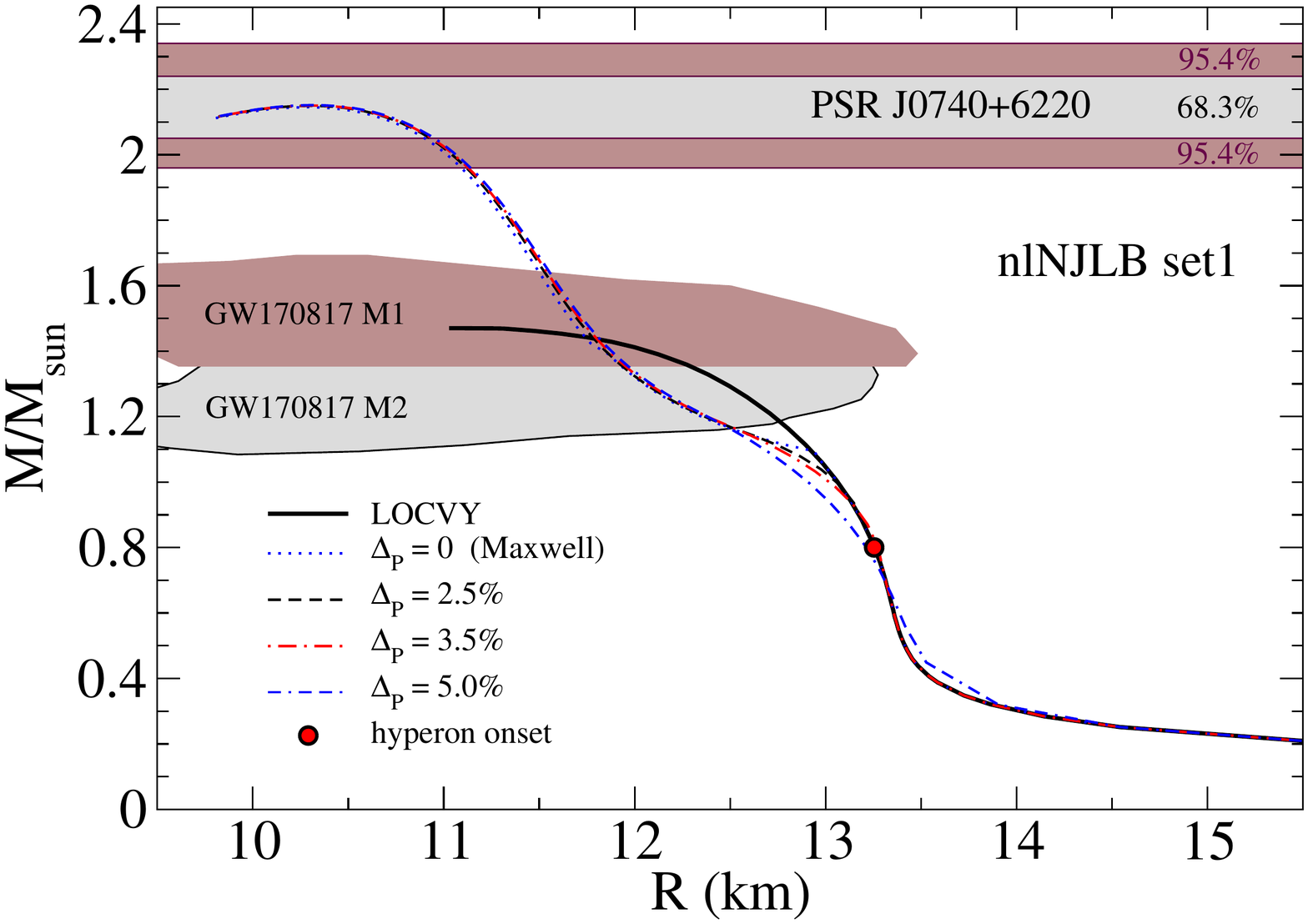}
\caption{\label{fig:5} 
Same as Fig. 4, but for hybrid stars based on the nlNJLB model of quark matter.}
\end{figure*}
The TOV equations have been solved for the EoS of hybrid stars which have been derived in previous sections and the mass-radius relation for them is shown in Fig. \ref{fig:4} and Fig. \ref{fig:5}. As Fig. \ref{fig:5} shows, for $\Delta_P$ = 5\%, unlike our expectation, the deconfinement and hyperon onset almost happen at the same density. Therefore, the best values of $\Delta_P$ for model nlNJLB are below 5\%. 
Fig. \ref{fig:4} shows the results of a hybrid star sequence with mixed phase construction for model nlNJLA while the results for model nlNJLB are shown in 
Fig. \ref{fig:5}.

By employing a mixed phase construction, the nlNJLA model now predicts a hybrid star with quark matter core surrounded by a hypernuclear matter shell before the transition to nuclear matter happens, unlike in paper I where a Maxwell construction was imposed and the nuclear matter would deconfine before the hyperon onset was reached. Thus, this nonphysical transition at very low densities (see the dotted line in Fig.~\ref{fig:4}) is ignored because the quark matter EoS is not valid in this region.
Moreover, in the present case with the mixed phase constructions, the mass-radius relations for both nlNJLA and nlNJLB models have a similar shape, without the branch with a positive slope that was obtained for the Maxwell construction.

Furthermore, the obtained maximum masses obey the current NS constraints:
$M_{\rm max} > 2.07~M_\odot$ from the lower limit of the $1\sigma$ range of the Shapiro-delay-based mass measurement for the millisecond pulsar PSR J0740+6620 \cite{Cromartie:2019kug}, as well as the lower and upper limits for the radius, $R_{1.6M_\odot}>10.7$ km \cite{Bauswein:2017vtn} 
and $R_{1.4M_\odot}<13.6$ km \cite{Annala:2017llu}, respectively,  from the binary NS merger GW170817 \cite{TheLIGOScientific:2017qsa}.

Another constraint which has to be fulfilled by the EoS of the hybrid star is the tidal deformability of GW170817. 
The quantity $\Lambda$ is the tidal deformability which is the induced quadrupole polarizability. 
The dimensionless tidal deformability $\Lambda$ is defined as \cite{TheLIGOScientific:2017qsa}
\begin{equation}
  \Lambda = \frac{2}{3}k_2\left(\frac{c^2R}{GM}\right)^5 = \frac{2}{3}k_2\left(\frac{1}{C}\right)^5,
   \label{tidal}
  \end{equation}
where $c$ is the speed of light, $R$ and $M$ are the radius and mass of the NS and $G$ is the gravitational constant. 
The compactness of the NS is presented by $C$ which is defined as $C = \frac{GM}{c^2R}$. $k_2$ is the tidal Love number which describes the response of each star to the external disturbance and depends on the structure of the NS and therefore on it's mass and EoS. 
It is clear that $\Lambda$ is very sensitive to the compactness parameter $C$ and also proportional to $k_2$. $k_2$ is also related to $C$ and $y_R$ which is a dimensionless parameter sensitive to the EoS \cite{Hinderer:2007mb, Hinderer:2009ca}.

The binary NS merger GW170817 could only acknowledge the upper limit on the tidal deformability of a $1.4M_\odot$ NS, i.e., $\Lambda_{1.4} \leq 800$ \cite{Piekarewicz:2018sgy} extracted from the original discovery paper \cite{TheLIGOScientific:2017qsa}. 
For comparison, the tidal deformability and it's parameters have been calculated for all hybrid stars predicted by mixed phase construction for $ M = 1.4M_{\odot} $ and the results are given in table \ref{tab:tidal}.
\begin{table}[h]
 \centering
 \caption{\label{tab:tidal}
  Central density $ \rho_{c} $ [fm$^{-3}$], radius $R$ [km], compactness $ C $, $ y_{R} $, tidal Love number $ k_{2} $ and dimensionless tidal deformability $ \Lambda $ for the hybrid stars with the mass of $ 1.4M_{\odot} $ using LOCVY method for the hadronic phase and the nlNJLA or the nlNJLB model for the quark phase with different mixed phase parameters 
  $\Delta_P$.}
  \vspace{5mm}
 \begin{tabular}{lcccccccc}
 model&$\Delta_P$ & $\rho_{c}$ & $R$ & $C$ & $y_R$ & $k_2$ & $\Lambda$ \\
 \hline
\textrm{nlNJLA}&-0.025 & 0.4413 & 12.50 & 0.1651 & 0.3560 & 0.09799   & 532.25 \\
 \textrm{nlNJLA}&-0.035 & 0.4399 & 12.59 & 0.16397 & 0.3534 & 0.09892   & 556.23  \\
 \textrm{nlNJLA}&-0.050 & 0.4291 & 12.69 & 0.1626 & 0.3527 & 0.09982  & 583.95  \\ \hline
 \textrm{nlNJLB}&0.025 & 0.570 & 11.843  & 0.17444 & 0.3611 & 0.0917 & 378.85  \\
 \textrm{nlNJLB}&0.035 & 0.570 & 11.849 & 0.1743 & 0.3597 & 0.0918   & 380.82  \\
 \textrm{nlNJLB}&0.050 & 0.568 & 11.86 & 0.174 & 0.3596 & 0.0920   & 384.23  \\
 %
 \end{tabular}
 \end{table}
 For the nlNJLA EoS we find the higher values of tidal deformability in comparison to the nlNJLB EoS. 
 The reason is that the mass of $ 1.4M_{\odot} $ occurs in the mixed phase region with the higher radii.
 Therefore, the star becomes less compact and from $ \Lambda=\frac{2}{3}k_{2}(\frac{c^{2}R}{GM})^{5}$, $ C=\frac{GM}{c^{2}R} $, the tidal deformability is proportional to $(\frac{1}{C})^{5}$.
But for the nlNJLB EoS, the mass of $ 1.4M_{\odot} $ occurs in the mixed phase region with lower radii and so the star is much more compact. 
Therefore, we get lower values of tidal deformability.
The value of tidal deformability and hence the radii of hybrid stars are compatible with the constraint on the hybrid star, 
$ 35.5<\tilde{\Lambda}_{1.4}<800 $ and $ 8.35 $ km $<R_{1.4}$   $ <13.74  $ km and the constraint for tidal deformability of neutron stars,  
which is $ 375<\tilde{\Lambda}_{1.4}<800 $ and  $ 12.00 $ km $<R_{1.4}$ $<13.45 $ km \cite{Piekarewicz:2018sgy}.
\section{Conclusion} 
 In this paper we have applied for the first time the finite-range polynomial interpolation method for constructing a transition between hadronic and quark matter phases to the situation that is characterized in the literature as the reconfinement problem.
Employing such a method for constructing a deconfinement phase transition from the 
hadronic phase of hypernuclear matter to the deconfined quark matter phase could mimics the effects of pasta structures in the mixed phase. 
It was illustrated that applying this method to both types of hybrid stars which have been obtained using nlNJLA and nlNJLB, one can improve the results such 
that for both cases, a mixed phase region is predicted between nuclear matter and quark matter phases which consists of modified hypernuclear and deconfined quark matter subphases.

It has been shown that all predicted hybrid stars using the nlNJLA and nlNJLB model could solve the hyperon puzzle and fulfill all recent constraints on the mass and radius of NS. 
For both types of EoS, obtained from nlNJLA and nlNJLB, the shape of the mass-radius curves is now similar. 
The dimensionless tidal deformability of predicted hybrid stars with $M=1.4M_\odot$ has been calculated and is in agreement with the value extracted from the observed gravitational waves from the inspiral phase of GW170817.

The application of the quark-hadron hybrid EoS to binary merger simulations, however, requires the inclusion of finite temperatures which is planned for the future extension of the presented approach. 

\subsection*{Acknowledgements}
We acknowledge Gabriela Grunfeld for providing data from the nonlocal chiral quark model that have been used in the present work
and Anthony Thomas for a comment on the hyperon puzzle.
M.S. is grateful to the research council of the University of Tehran for supporting her research visit at the University of 
Wroclaw where this work has been performed, and to the HISS Dubna program for supporting her participation in the 
summer schools in 2018 and 2019 where this project was initiated and completed, also with the support of the 
Bogoliubov-Infeld program for scientist exchange between Polish Institutes and JINR Dubna. 
The work of D.B. was supported by 
the Polish National Science Center (NCN) under grant no. UMO-2019/33/B/ST9/03059 and by the National Research 
Nuclear University (MEPhI) within the Russian Academic Excellence Project under contract no. 02.a03.21.0005.
D.B and M.S. thank the European COST Actions CA15213 "THOR" and  CA16214 "PHAROS" for supporting their networking activities.


\end{document}